\documentclass[10pt,twocolumn]{article} 
\usepackage{simpleConference}
\usepackage{cite}
\usepackage{amsmath,amssymb,amsfonts}
\usepackage{algorithmic}
\usepackage{graphicx}
\usepackage{textcomp}
\usepackage{subfig}
\usepackage{textcomp}
\usepackage{stfloats}
\usepackage{url}
\usepackage{verbatim}
\usepackage{cite}
\usepackage{amsmath,adjustbox}
\usepackage[table,xcdraw]{xcolor}
\usepackage{graphicx,xcolor} 
\usepackage{booktabs}
\usepackage[table,xcdraw]{xcolor}
\usepackage{caption}
\usepackage{hyperref}

\begin{document}
	\title{Lateral Strain Imaging using Self-supervised and Physically Inspired Constraints in Unsupervised Regularized Elastography}
	\author{Ali~K. Z. Tehrani, Md Ashikuzzaman, and Hassan~Rivaz 
		\thanks{A. K. Z. Tehrani,  Md Ashikuzzaman, and H. Rivaz are with the Department
			of Electrical and Computer Engineering, Concordia University, Canada,
			e-mail: A\_Kafaei@encs.concordia.ca,
			m\_ashiku@encs.concordia.ca and
			hrivaz@ece.concordia.ca
		}
		\thanks{}}
	
	\maketitle
	
	\begin{abstract}
		Convolutional Neural Networks (CNN) have shown promising results for displacement estimation in UltraSound Elastography (USE). Many modifications have been proposed to improve the displacement estimation of CNNs for USE in the axial direction. However, the lateral strain, which is essential in several downstream tasks such as the inverse problem of elasticity imaging, remains a challenge. The lateral strain estimation is complicated since the motion and the sampling frequency in this direction are substantially lower than the axial one, and a lack of carrier signal in this direction. In computer vision applications, the axial and the lateral motions are independent. In contrast, the tissue motion pattern in USE is governed by laws of physics which link the axial and lateral displacements. In this paper, inspired by Hooke's law, we first propose Physically Inspired ConsTraint for Unsupervised Regularized Elastography (PICTURE), where we impose a constraint on the Effective Poisson’s ratio (EPR) to improve the lateral strain estimation. In the next step, we propose self-supervised PICTURE (sPICTURE) to further enhance the strain image estimation. Extensive experiments on simulation, experimental phantom and \textit{in vivo} data demonstrate that the proposed methods estimate accurate axial and lateral strain maps.
	\end{abstract}
	

	\section{Introduction}
	Ultrasound (US) imaging is a popular modality in diagnosis and image-guided interventions thanks to its portability, affordability, and non-invasiveness. Ultrasound Elastography (USE) is an imaging technique that provides information about the stiffness of the tissue. An external or internal force is applied to compress the tissue, and US images before and after deformation are analyzed to find the displacement map~\cite{ophir1999elastography,varghese2000direct}. In free-hand palpation USE, the external force is a compression applied by the operator using the probe. The tissue mostly compresses in the axial direction while expanding in the other two directions as well. The axial and lateral strain maps, which are obtained by taking derivatives of the displacements can be used as a surrogate of elastic properties of tissues. USE has been found clinically useful, especially when the B-mode images do not clearly discern the target tissue. As such, USE has been employed in different clinical applications, including monitoring of thermal ablation \cite{kling2018potential,lee2019evaluating}, assessment of thyroid gland tumors \cite{Lyshchik2005}, and characterization of breast lesions \cite{hall2001vivo}. Conventional USE methods find the displacements using block matching (window-based techniques) \cite{varghese2000direct,nahiyan2015hybrid,luo2010fast,jiang2007parallelizable,mirzaei20203d} or optimization strategy \cite{hashemi2017global,mirzaei2019combining, ashikuzzaman2021combining,fleming2014robust,mccormick2011bayesian}. 

	Deep learning methods have attracted growing attention due to their ability to learn complex mappings. Recently Convolutional Neural Networks (CNN) have been employed for USE \cite{Kibria2018,wu2018direct,gao2019learning,peng2018convolution,evain2020pilot,peng2020neural,tehrani2020real}, ultrasound frame selection for USE \cite{zayed2020fast}, elasticity reconstruction \cite{mohammadi2021combining}, acoustic radiation force imaging \cite{chan2021deep}, and echocardiography motion estimation \cite{evain2022motion}. The architecture of the networks was modified to adapt the network to USE \cite{tehrani2020displacement,tehrani2021mpwc}. MPWC-Net++, designed based on the well-known PWC-Net architecture \cite{sun2018pwc,hur2019iterative}, was able to handle radio-frequency (RF) data as inputs. The correlation search range of PWC-Net was increased to enable the network to handle large image sizes, and the downsampling scale of the feature extraction block of the network was decreased from 4 to 2 to avoid loss of high-frequency RF data \cite{tehrani2021mpwc} information. The weights were made available online at \href{http://code.sonography.ai}{code.sonography.ai}. 
	
	Unsupervised training of USE network is another avenue that has been followed to enhance the displacement estimation in USE. Unsupervised training on real ultrasound data prepares the trained model to extract more suitable features from RF data \cite{delaunay2020unsupervised,delaunay2021unsupervised,tehrani2020semi,tehrani2022bi,wei2022unsupervised}. Prior knowledge of displacement map continuity is also utilized in the forms of different regularization strategies. In \cite{tehrani2020semi}, a combination of first and second-order derivatives of the displacements is employed as the regularization, which has been found beneficial in the recent optimization-based methods \cite{ashikuzzaman2021combining,ashikuzzaman2022second}. Wei \textit{et al.} adapted MaskFlownet \cite{zhao2020maskflownet} to USE and trained the network using an unsupervised method. They also performed a detailed comparison of their network with MPWC-Net++ \cite{wei2022unsupervised}.     

	In classical methods, Babaniyi \textit{et al.} \cite{babaniyi2017recovering} considered plane stress and incompressibility assumptions to refine the estimated displacement. Guo \textit{et al.} first introduced a refinement method that incorporated the incompressibility and plane strain assumptions in an iterative approach \cite{guo2015pde} that substantially improved the lateral strain. Other lateral strain imaging works mainly focus on modifying the imaging technique to have a higher resolution in lateral direction \cite{wang2022hadamard,mirzaei2020virtual,selladurai2018strategies}, and, as such, cannot be applied to the already available US data. The smoothness of the derivatives of the displacements is the only prior knowledge of USE physics used in previous unsupervised training. No deep learning work considers the physics of the compression of the tissue into account. Also, no deep learning method has focused on improving the quality of the lateral displacement estimation, which is challenging but it is required for elasticity \cite{mohammadi2021ultrasound} and Poisson’s ratio imaging \cite{islam2018new}.  
	
	In our preliminary work, we investigated the feasibility of improving the lateral displacement by employing the prior knowledge of compression physics \cite{kz2022physically}, where we introduced Physically Inspired ConsTraint for Unsupervised Regularized Elastography (PICTURE). In this paper, the method is explained in more detail, and new extensive experiments are performed to better evaluate the effectiveness of the technique. We also introduce self-supervision in USE and propose sPICTURE, which further boosts the performance.

	\section{Method}
	
	\subsection{Unsupervised Training}
	Let  $I_1,I_2\in \mathbb{R}^{3\times W \times H}$ denote the pre- and post-compression US data, respectively. The subscript $3$ refers to three channels of RF data, the envelope of RF data, and the imaginary part of the analytic signal similar to \cite{tehrani2020real}. The unsupervised cost function is composed of data loss and smoothness regularization loss. The data loss ($L_D$) in unsupervised training can be obtained by comparing $I_1$ with the warped $I_2$ ($\hat{I_2}$) by the displacement map $W$. The data loss can be written as \cite{meister2018unflow,tehrani2022bi}:
	\begin{equation}
		\label{eq:loss_d}
		L_D =  ||(I_1-\hat{I_2})||_{1(N\times N)} 
	\end{equation}  
	where $||.||_1$ denotes norm 1 (as suggested by \cite{sun2018pwc,hur2019iterative}, L2 norm is not suitable; therefore, a norm lower is employed), and a window of size $N \times N$ (here $3 \times 3$) is considered around each sample to reduce the effect of noise. For the regularization, we adopt the method of \cite{tehrani2022bi,tehrani2020semi} where the strains and their first-order derivatives are employed. The strains can be obtained by taking the derivative of displacement in direction ($x$) with respect to the direction ($y$):
	\begin{equation}
		\begin{gathered}	
			\varepsilon_{xy} = \frac{\partial W_x}{\partial y}\\
			x,y \in 1, 2, 3
		\end{gathered}	
	\end{equation}
	we assumed that the subscripts 1, 2, and 3 denote axial, lateral, and out-of-plane directions, respectively. By this assumption, $\varepsilon_{11}$, $\varepsilon_{22}$ and, $(\varepsilon_{21}+\varepsilon_{12})/2$ are the axial, lateral and, shear strains, respectively. The smoothness loss can be defined as:     
	
	\begin{equation}
		\label{eq:loss_s}
		\begin{gathered}	
			L_S = L_{s1} + \gamma L_{s2}\\
			L_{s1} = ||\varepsilon_{11}-<\varepsilon_{11}>||_1 + \beta||\varepsilon_{12}||_1+\frac{1}{2}||\varepsilon_{21}||_1+\frac{1}{2}\beta||\varepsilon_{22}||_1 \\	
			L_{s2} = \biggl \{||\ (\frac{\partial \varepsilon_{11}}{\partial a})||_1 + \beta ||\ (\frac{\partial \varepsilon_{11}}{\partial l})\biggr.||_1 + \\ \left. 0.5||\ (\frac{\partial \varepsilon_{22}}{\partial{a}})||_1+0.5\beta||\ (\frac{\partial \varepsilon_{22}}{\partial l}) ||_1\right \}
		\end{gathered}
	\end{equation}

	where $L_S$ is the total smoothness loss, $<.>$ denotes averaging operation, and $\beta$ depends on the ratio of spatial distance between two samples in lateral to the axial direction and it is set to 0.1 similar to \cite{tehrani2022bi}. $L_{s1}$ and $L_{s2}$ are the regularization of first- and second-order derivatives 
	of the displacements.  $\gamma$ is a hyperparameter that controls the weight of the second-order derivatives smoothness loss. 
	\subsection{Hooke's Law and compression physics}
	Assuming that the tissue is linear elastic and isotropic, the following two sections show the relation between the lateral and axial displacements under uniform compressions.
	\subsubsection{Homogeneous Material}  
	Hooke's law can be formulated as \cite{ugural2003advanced}:
	
	\begin{equation}
		\label{eq:hook}
		\resizebox{0.48\textwidth}{!}{$
			\begin{bmatrix}
				\varepsilon_{11}\\ 
				\varepsilon_{22}\\ 
				\varepsilon_{33}\\ 
				2\varepsilon_{23}\\ 
				2\varepsilon_{13}\\ 
				2\varepsilon_{12}
			\end{bmatrix}
			= \frac{1}{E}
			\begin{bmatrix}
				1 & -v & -v & 0 & 0 & 0\\ 
				-v& 1 & -v & 0 & 0 & 0\\ 
				-v&  -v&  1&  0&  0& 0\\ 
				0&  0&  0&  2+2v&  0& 0\\ 
				0&  0&  0&  0&  2+2v& 0\\ 
				0&  0&  0&  0&  0&2+2v 
			\end{bmatrix}
			\begin{bmatrix}
				\sigma_{11}\\ 
				\sigma_{22}\\ 
				\sigma_{33}\\ 
				\sigma_{23}\\ 
				\sigma_{13}\\ 
				\sigma_{12}
			\end{bmatrix}$}
	\end{equation}
	where $E$, $\sigma$, and $v$ represent Young's modulus, stress tensors, and Poisson's ratio, respectively. When there is a compression of the material in one direction, there is an expansion in the other direction, which depends on the Poisson’s ratio of the material. In free-hand palpation, it can be assumed that the external force is only in the axial direction (uniaxial stress); therefore, other stress components except $\sigma_{11}$ can be ignored. This assumption simplifies Eq \ref{eq:hook} and leads to \cite{ugural2003advanced}:
	\begin{equation}      
		\varepsilon_{11} = \frac{\sigma_{11}}{E}, \varepsilon_{22} = -v\frac{\sigma_{11}}{E},  \varepsilon_{33} = -v\frac{\sigma_{11}}{E} 
	\end{equation}
	which indicates that the lateral strain ($\varepsilon_{22}$) can be directly obtained by the axial one ($\varepsilon_{11}$) using $-v \times \varepsilon_{11}$. 
	\subsubsection{Inhomogeneous Materials}
\begin{figure}[!t]
		\centering
		\includegraphics[width=0.49\textwidth]{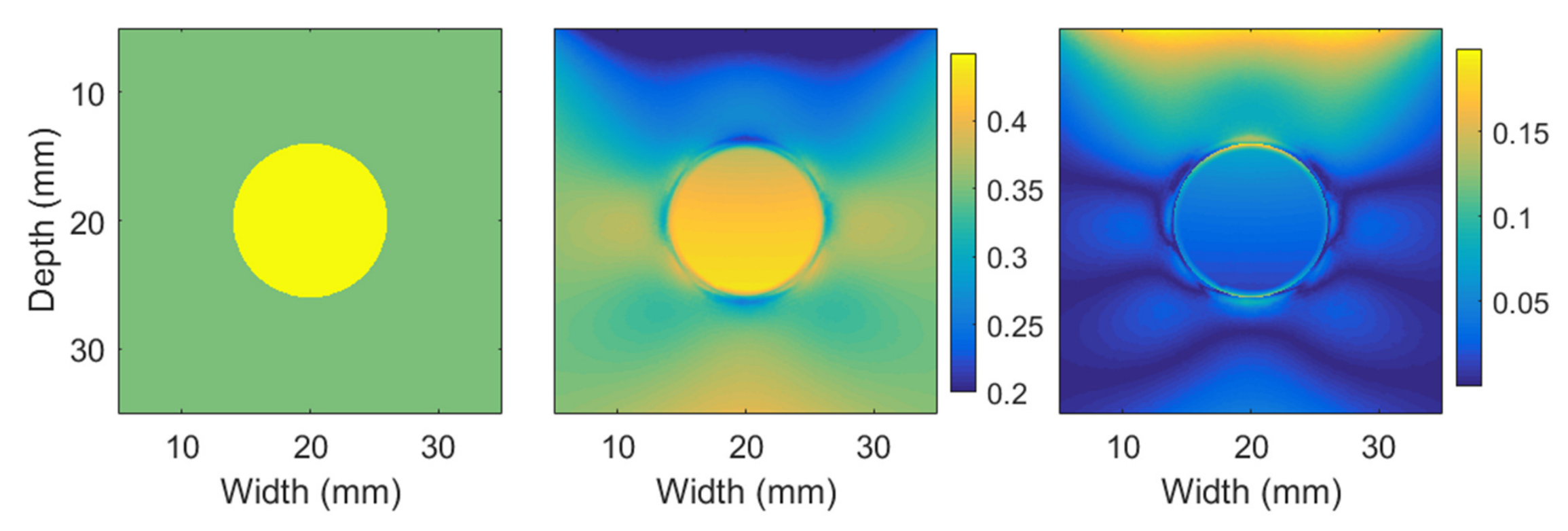}
		\centering
		\caption{From left to right: the Poisson’s ratio, the EPR, and their absolute difference for a simulated phantom. The Poisson’s ratio and the EPR have the same colorbar. } 
		\label{fig:EPR}
	\end{figure}

	Tissues cannot be assumed to be homogeneous due to the presence of irregularities and boundary regions; therefore, the lateral strain cannot be directly obtained by the axial one and Poisson’s ratio. In this condition, the total strain ($\varepsilon_{xy}$) is obtained by adding the elastic strain ($e_{xy}$) and eigenstrain ($\varepsilon_{xy}^{\ast}$) \cite{ma2014principle}:
	\begin{equation}
		\varepsilon_{xy} = 	e_{xy} + \varepsilon_{xy}^{\ast}
	\end{equation}
	Eigenstrain is added to consider the variation of total strain from the elastic one in the presence of inhomogeneity. It is maximum on the inhomogeneity boundaries and decays to zero further from the boundaries \cite{ma2014principle}. Although the lateral strain does not linearly depend on the axial one anymore, they are still highly correlated. Also, $-\varepsilon_{22}/\varepsilon_{11}$ does not obtain the Poisson’s ratio anymore and it is called Effective Poisson’s ratio (EPR) \cite{islam2018new}. In uniform regions far from inhomogeneities, EPR converges to Poisson’s ratio. For illustration purpose, EPR and Poisson’s ratio of a finite element simulation using ABAQUS software (Providence, RI) is depicted in Fig. \ref{fig:EPR}. It can be observed that EPR is more dissimilar to Poisson’s ratio at the top and bottom of the phantom and around the inclusion. Poisson’s ratio and EPR under arbitrary deformation have the range between 0.2 and 0.5 \cite{mott2013limits,righetti2004feasibility}. Although the exact value of EPR is not known, it has been used as an approximation of Poisson’s ratio to characterize tissues \cite{islam2018new,righetti2004feasibility}. We propose to use this range as a prior information to guide the network to refine the lateral displacement. Guo \textit{et al.} assumed tissue incompressibility (Poisson's ratio = 0.5) and plane strain (strain in out-of-plane direction = 0) to refine the displacements. However, Poisson's ratio depends on the type of the tissue (refer to \cite{george2018influence,griesenauer2017breast} for Poisson's ratio of breast and liver). In this work, we do not make those assumptions, and only a feasible range of Poisson's ratio is enforced.
	
	\subsection{PICTURE}
	We propose to utilize the accepted range of EPR as a prior information in the form of regularization. To do that, we first need to calculate EPR ($V_e$) from the estimated axial and lateral strains:
	\begin{equation}
		V_e = \frac{-\tilde{\varepsilon_{22}} }{\mathbf{S}(\tilde{\varepsilon_{11}})} 
	\end{equation}
	where $\tilde{\varepsilon_{22}}$ and $\tilde{\varepsilon_{11}}$ are the lateral and axial strains obtained from estimated displacements. The parameter $\mathbf{S}$ denotes stop gradient operation, which is used to stop backpropagation of the loss to the axial displacement estimation. It is used to avoid the estimated axial displacement being altered by the noisy lateral one. To find out the EPR values outside the accepted range, a mask ($M$) is defined using the minimum ($V_{emin}$) and maximum  ($V_{emax}$) allowed EPR values.
	\begin{equation}
		M(i,j) = \begin{Bmatrix}
			0 &  & V_{emin}<V_e(i,j)<V_{emax} \\
			1&  &  otherwise\\
		\end{Bmatrix}
	\end{equation}    
	We assume the $V_{emin}$ and $V_{emax}$ values to be 0.1 and 0.6 (no noticeable change was observed by small changes of these values) to have a small margin of error. In order to penalize the EPR values outside of the accepted range, PICTURE loss is defined as:
	\begin{equation}
		\label{Eq:picture_loss}
		\begin{gathered}
			L_{vd} =\left|\left| M\otimes (\tilde{\varepsilon_{22}}+\bar{V_e}\times \mathbf{S}(\tilde{\varepsilon_{11}}))\right|\right|_2 
		\end{gathered}
	\end{equation} 
	where $\otimes$ is the Kronecker product to select EPR values outside the accepted range, and $\bar{V_e}$ is an estimate of true average EPR. It is obtained by averaging EPR values that are inside the accepted range, which can be formulated as:
	\begin{equation}
		\bar{V_e} = \frac{\sum_{i,j}^{}(1-M_{(i,j)}) V_e(i,j)}{\sum_{i,j}^{}(1-M_{(i,j)})}
	\end{equation}  
	Eq \ref{Eq:picture_loss} tries to constrain EPR to be inside the accepted range.

	The first-order derivatives of $V_e$ are also employed to enforce the smoothness of EPR.
	\begin{equation}
		L_{vs} = ||\frac{\partial V_e}{\partial a}||_1 + \beta \times ||\frac{\partial V_e}{\partial l}||_1
	\end{equation}  
	The final PICTURE loss can be written as:
	\begin{equation} 
		L_V = L_{vd} + \lambda_{vs} \times L_{vs}
	\end{equation} 
	where $\lambda_{vs}$ is defined to weight the smoothness part.
	\subsection{Self-Supervised Learning}
	Self-supervised learning (SSL) is a technique that has recently been applied to unsupervised optical flow networks \cite{liu2020learning,liu2019selflow}. The basic procedure is that the input images are fed to the network during the unsupervised training, and the displacements are obtained during the \textit{first pass}. In the next step, the input images are transformed to make them more challenging than before and the new displacement is obtained during the \textit{second pass}. In the last step, the differences between the displacements of the first and second passes are penalized:
	\begin{equation}
		\label{eq:ssl} 
		L_{SSL} = ||\mathbf{S}(W)-\tilde{W}||_1
	\end{equation}
	where $W$ is the estimated displacement in the first pass (no transformation), $\tilde{W}$ is the obtained displacement from the second pass (with transformed inputs), and stop gradient ($\mathbf{S}$) is used to avoid backpropagation into the first pass. 
	Substantial improvements were reported for unsupervised training employing SSL using different transformations. In \cite{liu2019selflow}, superpixels \cite{ren2003learning} of input images were identified and the content of randomly selected superpixels were replaced by pure noise. The method outperformed other unsupervised methods in different optical flow benchmarks. In \cite{liu2020learning}, cropping, affine, and other kinds of transformations were utilized. SSL was also compared by data augmentation (instead of SSL, the transformed images were considered as new inputs). SSL outperformed data augmentation, which indicated that SSL was not a simple synthetic data generation like data augmentation.  
	\begin{figure}[!t]
		\centering
		\includegraphics[width=0.44\textwidth]{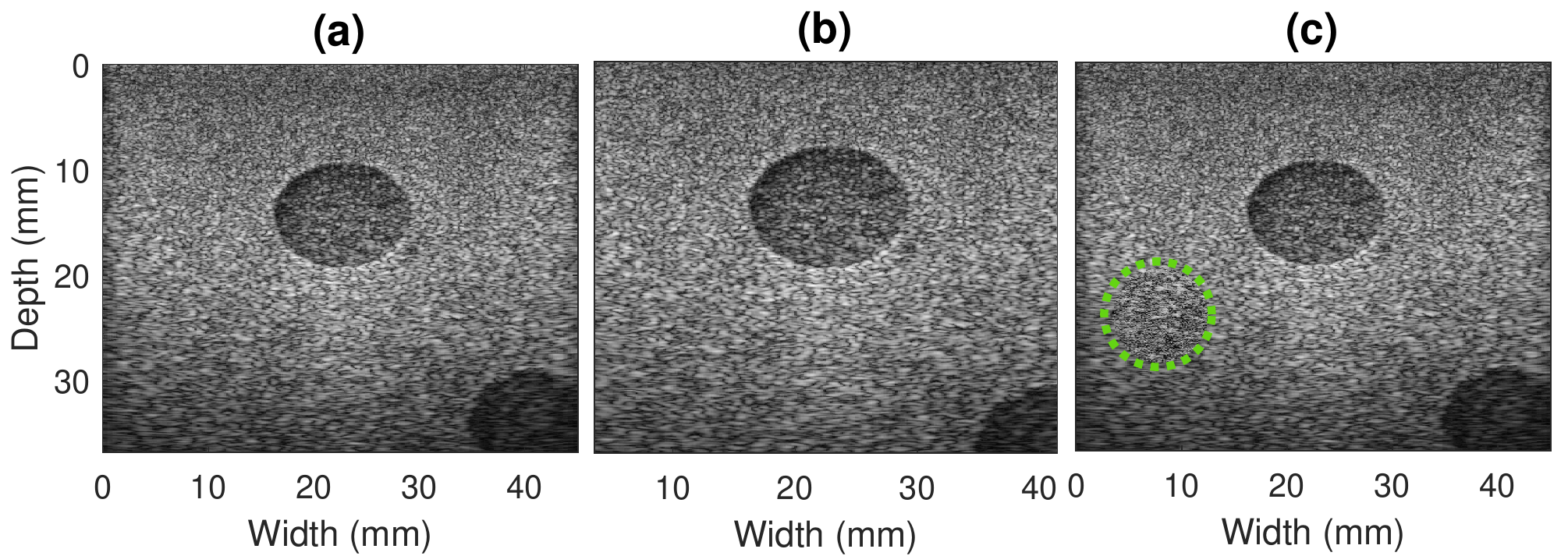}
		\centering
		\caption{B-mode input image (a), cropped image (resized for the purpose of visualization) (b) and, image with added noise (the area of added noise is highlighted) (c).} 
		\label{fig:ssl}
	\end{figure}
	
	In this paper, we employ two data transformations: cropping and adding noise to  specific regions. Cropping may cause loss of information, especially in areas where displacement is high. On those areas, estimating displacement is complex for the network since the corresponding part of the first image might be outside the cropped second image. We also add large Gaussian noise to randomly selected circular regions. An example of a transformed image is shown in Fig. \ref{fig:ssl}. SSL can guide the network to have a more reliable estimation when there is a loss of information due to cropping or noisy data. 
	
	\subsection{Loss Function And Network Architecture}
	The loss function is composed of data loss (Eq \ref{eq:loss_d}), smoothness loss (Eq \ref{eq:loss_s}), PICTURE loss (Eq \ref{Eq:picture_loss}), and SSL regularization (Eq \ref{eq:ssl}):
	\begin{equation}
		\label{eq:loss_total}
		loss = \underbrace{L_D + \lambda_s L_S + \lambda_v L_V}_{first \ pass} + \underbrace{\lambda_{sl}L_{SSL}}_{second \ pass}
	\end{equation}
	where the hyper-parameters, $\lambda_s$, $\lambda_v$, and $\lambda_{sl}$ are the weights of smoothness regularization, PICTURE loss, and SSL regularization, respectively. The SSL loss only affects the second pass in which the input US data are transformed, while the other losses affect the first pass.  
	
	We employed MPWC-Net++ \cite{tehrani2021mpwc} as the network. This network has demonstrated promising results on USE by increasing the search range of correlation layers and reducing the downsampling of PWC-Net-IRR \cite{hur2019iterative} feature extraction layers. The modification necessitated training the network from scratch; therefore, the FlyingChairsOcc dataset \cite{hur2019iterative} with 23000 image pairs was employed for the training. Refer to \cite{tehrani2021mpwc} for more details about the architecture of this network. The network's weights are publicly available online at \href{http://code.sonography.ai}{code.sonography.ai}.
	
	The hyperparameters values used for the training of unsupervised method and sPICTURE are given in the Supplementary Materials. The networks are trained for 25 epochs, the learning rate is set to $5 \times 10^{-6}$, which is halved every 5 epochs.

	
	\begin{figure*}[t]
		\centering
		\includegraphics[width=0.8\textwidth]{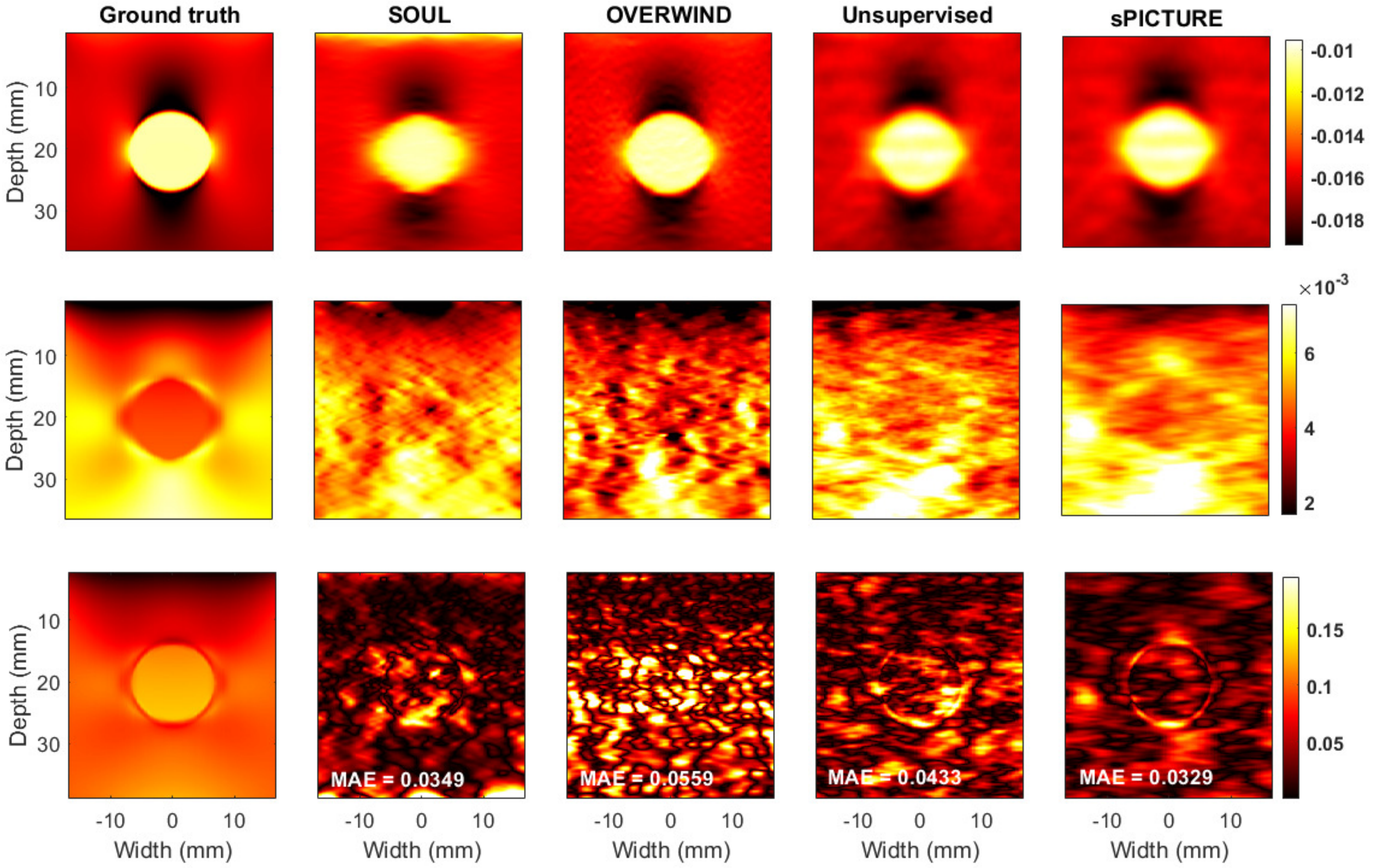}
		\centering
		\caption{Axial (top row) and lateral (middle row) strains in a simulated phantom. Ground truth EPR (bottom row), the absolute error, and the mean absolute error (MAE) shown for each method.}
		\label{fig:sim}
	\end{figure*}  
	
	\subsection{Datasets}
	\subsubsection{Simulation data}
	A phantom is simulated using Field II \cite{jensen1992calculation,jensen1996field}, and the motion is obtained by the ABAQUS finite element analysis software (Providence, RI). The phantom contains an inclusion with Poisson’s ratio of 0.45 and the Young's modulus of 40 kPa. The background has Poisson’s ratio of 0.35 and the Young's modulus of 21 kPa. The Poisson’s ratio and EPR of this phantom are shown in Fig. \ref{fig:EPR}. Different Poisson’s ratios for background and the inclusion are chosen to investigate if it is detectable by the networks. Compared to our recent simulation dataset \cite{tehrani2020displacement}, the number of lines in FIELD II is increased to 190, the number of active elements is increased to 96, and the obtained US images are also upsampled in the lateral direction by 2 to provide high lateral resolution.   
	
	In addition, 1200 pairs of publicly available simulated phantoms from \cite{tehrani2020displacement} are employed for training and 70 pairs for quantitative evaluation of the compared methods. These phantoms have a Poisson’s ratio of 0.49 and have one or two hard inclusions in different locations. The Young's modulus for the background and inclusions are around 20 kPa and 45-60 kPa, respectively.    
	
	\subsubsection{Experimental phantom data}
	RF data is collected from a tissue-mimicking breast phantom (Model 059, CIRS: Tissue Simulation \& Phantom Technology, Norfolk, VA) at Concordia University. It has Young's modulus of 20 kpa and contains several inclusions with the Young's modulus of at least 40 kPa. We employed an Alpinion E-Cube R12 research US machine (Bothell, WA, USA) for data collection. The sampling frequency was 40 MHz and the center frequency was 8 MHz. We made this data publicly available online at \href{http://code.sonography.ai}{code.sonography.ai} in \cite{tehrani2022bi}.
	
	2200 frame pairs are employed for training of the network. In order to avoid data leakage, different parts of the phantom were imaged for evaluation and test.
	
	\subsubsection{in vivo data}
	Data were acquired from patients with liver cancer during open-surgical RF thermal ablation at Johns Hopkins Hospital. A research Antares Siemens system by a VF 10-5 linear array was used for data collection. The sampling and center frequencies were 40 MHz and 6.67 MHz, respectively. The  study  was  approved  by  the  institutional  review board with the consent of all patients. 500 RF frame pairs from after ablation were selected for the training of the networks, and RF data from 2 patients before ablations were employed for test to prevent using similar data during the train and test phases.   
	
	\subsection{Quantitative Metrics}
	Mean Absolute Error (MAE) and Structural Similarity Index Metric (SSIM) \cite{Brunet2011} are employed for simulation results where the ground truth is available. For experimental phantoms, Contrast to Noise Ratio (CNR), and Strain Ratio (SR) are reported \cite{ophir1999elastography}:
	\begin{equation}
		\label{Eq:SRCNR}
		CNR = \sqrt{\frac{2(\overline{s}_{b}-\overline{s}_{t})^{2}}{{\sigma _{b}}^{2}+{\sigma _{t}}^{2}}},\quad \quad  SR =\frac{\overline{s}_{t}}{\overline{s}_{b}},
	\end{equation}
	where $\overline{s}_{X}$, and ${\sigma _{X}}$ are the mean and the standard deviation of obtained strain in the target (subscript $t$) and background (subscript $b$) windows. Assuming that the target is stiffer than the background, lower SR represents a higher difference between the average of strain in the target and background windows. In SR, the mean values of strains are employed; therefore, it is insensitive to the variance of strains. CNR combines both the mean and variance of the target and background windows which can provide a good intuition of the overall quality of the strain image.

	\begin{table*}[t]
		\caption{Quantitative results for 70 simulated phantoms. Mean and standard deviation ($\pm$) of the MAE of displacements and SSIM of strains are reported. The pairs marked by asterisk are not statistically significant (\textit{p}-value$>$0.05, using Friedman test).  }
		\centering
		\resizebox{0.70\textwidth}{!}{
			
			\begin{tabular}{ccccc}
				& \multicolumn{2}{c}{Axial} & \multicolumn{2}{c}{Lateral} \\ \cline{2-5} 
				& MAE $(\mu m)$        & SSIM $(\%)$       & MAE $(\mu m)$          & SSIM $(\%)$        \ \\ \hline
				\rowcolor[HTML]{C0C0C0} 
				SOUL& \textbf{2.2}$\pm$1.5 & \textbf{99.80}$\pm$0.06 & 8.00$\pm$4.1$^\ast$ & 97.70 $\pm$ 1.56$^\ast$\\
				OVERWIND     & \textbf{2.2}$\pm$1.5            & \textbf{99.80}$\pm$0.07            &9.40$\pm$4.6              &93.48$\pm$4.10              \ \\
				\rowcolor[HTML]{C0C0C0} 
				Unsupervised & 2.7$\pm$1.6$^\ast$            & 99.43 $\pm$ 2.10            & 8.70$\pm$4.1              &96.42$\pm$1.79              \ \\
				sPICTURE     & 2.7$\pm$1.7$^\ast$            & 99.55 $\pm$ 1.80           & \textbf{8.00}$\pm$3.8$^\ast$             & \textbf{97.73} $\pm$1.29$^\ast$              \  \\ \hline
		\end{tabular}}
		
	\end{table*}
	\begin{figure*}[]
		\centering
		\includegraphics[width=0.85\textwidth]{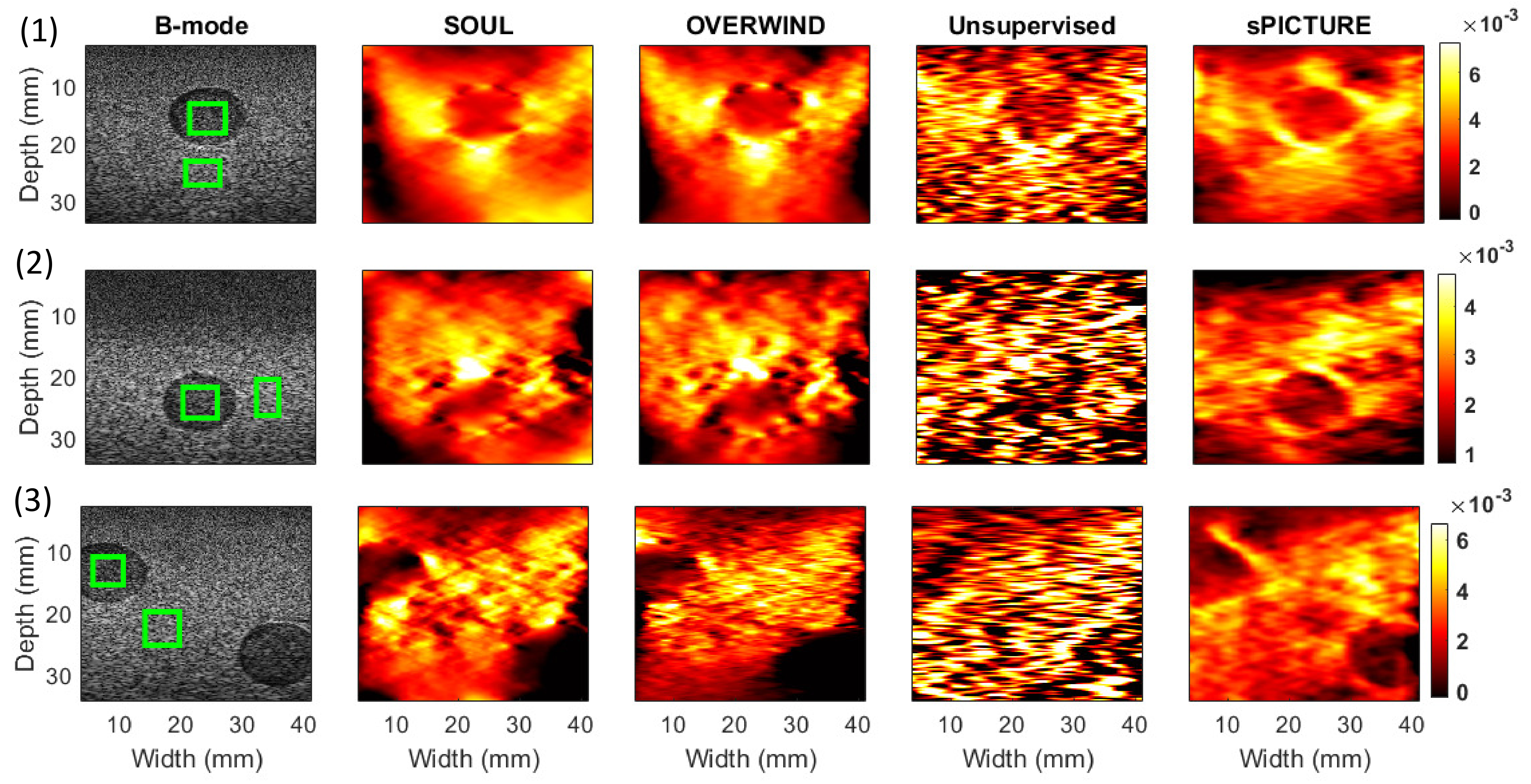}
		\centering
		\caption{The experimental phantom lateral strains obtained by the evaluated methods. The target and background windows for calculation of CNR and SR are marked in the B-mode images. The corresponding axial strains are shown in the supplementary video. The samples 1, 2 and 3 are taken from different locations of the tissue-mimicking breast phantom. Hard inclusions have lower absolute values than the background.} 
		\label{fig:phantom}
	\end{figure*}
	\section{Results}
	The evaluated methods are listed below:
	\begin{itemize}
		\item Second-Order Ultrasound Elastography (SOUL) is an optimization-based method which employs L2-norm and second-order regularization to have smooth strain images with high target-background contrast \cite{ashikuzzaman2021combining}.
		\item Total Variation Regularization and Window-based time delay estimation (OVERWIND) is a method that incorporate windowing into the optimization cost function \cite{mirzaei2019combining}.  
		\item Unsupervised method ($\lambda_{v} = 0$ and $\lambda_{sl} = 0$ in Eq \ref{eq:loss_total}, similar to the unsupervised method in \cite{tehrani2022bi}).
		\item The PICTURE without SSL ($\lambda_{sl} = 0$ in Eq \ref{eq:loss_total}, only used in ablation experiment).
		\item The proposed method named sPICTURE entails both PICTURE and SSL losses (Eq \ref{eq:loss_total}). 
	\end{itemize}
	It should be mentioned that for simulation results the network for unsupervised method and sPICTURE is trained on simulation data. For the experimental phantom results, it is trained on the experimental phantom dataset, and for \textit{in vivo} results, it is trained on the available \textit{in vivo} dataset. We also tuned the hyperparameters of the optimization-based methods (SOUL and OVERWIND) for each dataset separately to ensure that the best results are obtained from those methods.   
	
		

		\subsection{Simulation Results}
		The axial, lateral strains, and the EPR of the simulated phantom obtained by the compared methods are illustrated in Fig. \ref{fig:sim}. All methods obtain high-quality axial strains. The axial strain of the unsupervised method and sPICTURE are quite similar since PICTURE is only applied to the lateral displacement and keep the axial one untouched. Comparing the lateral strains (second row), sPICTURE reduces the noise presented in the unsupervised method. 
		
		The mean and standard deviations of quantitative metrics are reported for 70 simulated phantoms. Since the ground truth is available, MAE of displacement and SSIM of strain are reported for the axial and lateral displacements and strains. SOUL and OVERWIND have the lowest MAE error and highest SSIM in the axial direction. sPICTURE performs similar to the unsupervised method since PICTURE does not affect the axial direction. In lateral displacement estimation, sPICTURE reduces the lateral MAE of unsupervised method from 25.0 to 7.9, a decrease of more than three folds. It also outperforms the optimization-based methods in terms of MAE, with SSIM values close to those of SOUL.   
		
		\subsubsection{simulation results for different signal to noise ratios (SNR)}
\begin{figure}
			\centering
			\includegraphics[width=.49\linewidth]{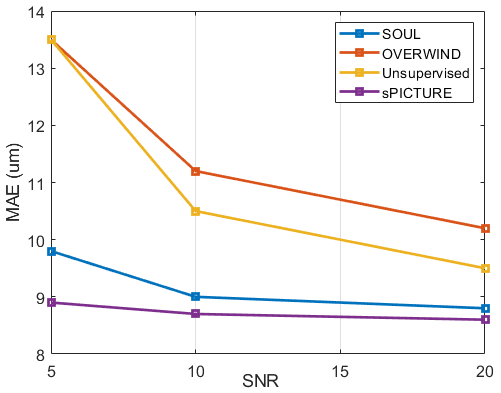} 
			\includegraphics[width=.49\linewidth]{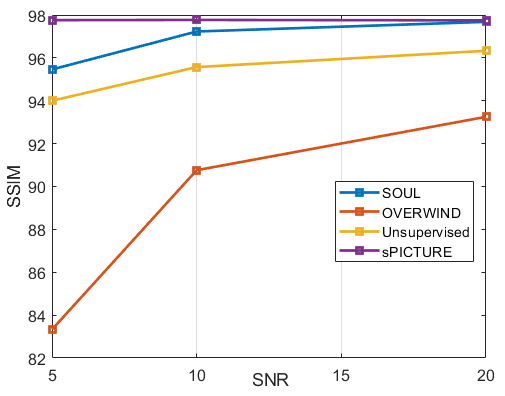}
			\caption{MAE of lateral displacements (left), and SSIM of lateral strains (right) for different SNR values of simulation test data.}
			\label{fig:snr}
		\end{figure}

		Random Gaussian noise with different SNRs is added to the test RF data to evaluate the robustness of the compared method to noise. MAE of lateral displacement and SSIM of lateral strain are plotted in Fig. \ref{fig:snr}. It can be observed that sPICTURE has a low MAE even for an SNR as low as 5 dB which demonstrates the high robustness of this method.    
		
		\subsubsection{simulation results for different compression}
\begin{figure}
			\centering
			\includegraphics[width=.7\linewidth]{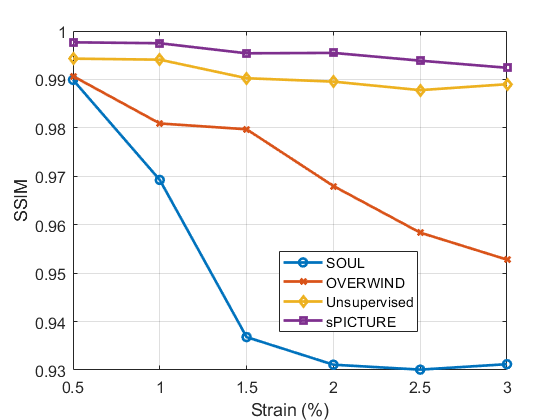} 
			\caption{SSIM of lateral strains versus different maximum strains.}
			\label{fig:strain}
		\end{figure}

		A phantom from simulation test data having different applied compressions, resulting in different maximum strains, is selected, and SSIM of lateral strain are illustrated in Fig. \ref{fig:strain}. By increasing the maximum strain, the SSIM of all compared method decreased which is expected. It should be noted that the graph shows that sPICTURE has the highest SSIM among the compared methods which is also robust to the variations of applied compression.              
		\subsection{Experimental Phantom Results}
		The lateral strains of experimental phantom results are shown in Fig \ref{fig:phantom} and the quantitative results are reported in Table \ref{tab:phantom}.

		\begin{figure}[]
			\centering
			\includegraphics[width=0.45\textwidth]{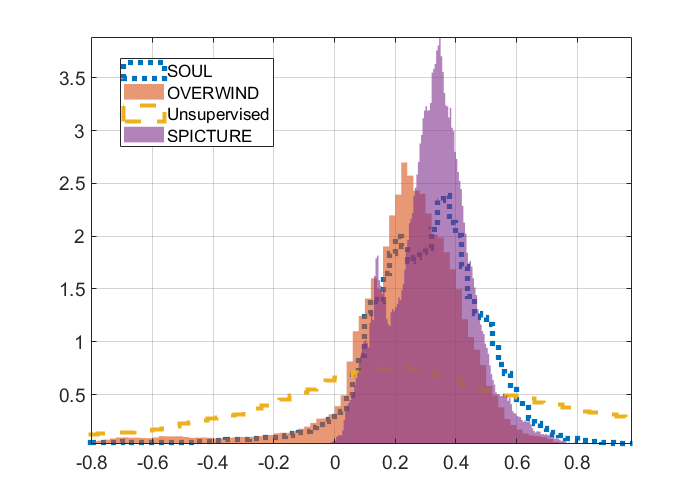}
			\centering
			\caption{The histograms of the EPR of different methods for phantom 3. sPICTURE has limited the EPR to the feasible range for USE.} 
			\label{fig:hist}
		\end{figure} 
		
		\begin{figure}[]
			\centering
			\includegraphics[width=0.48\textwidth]{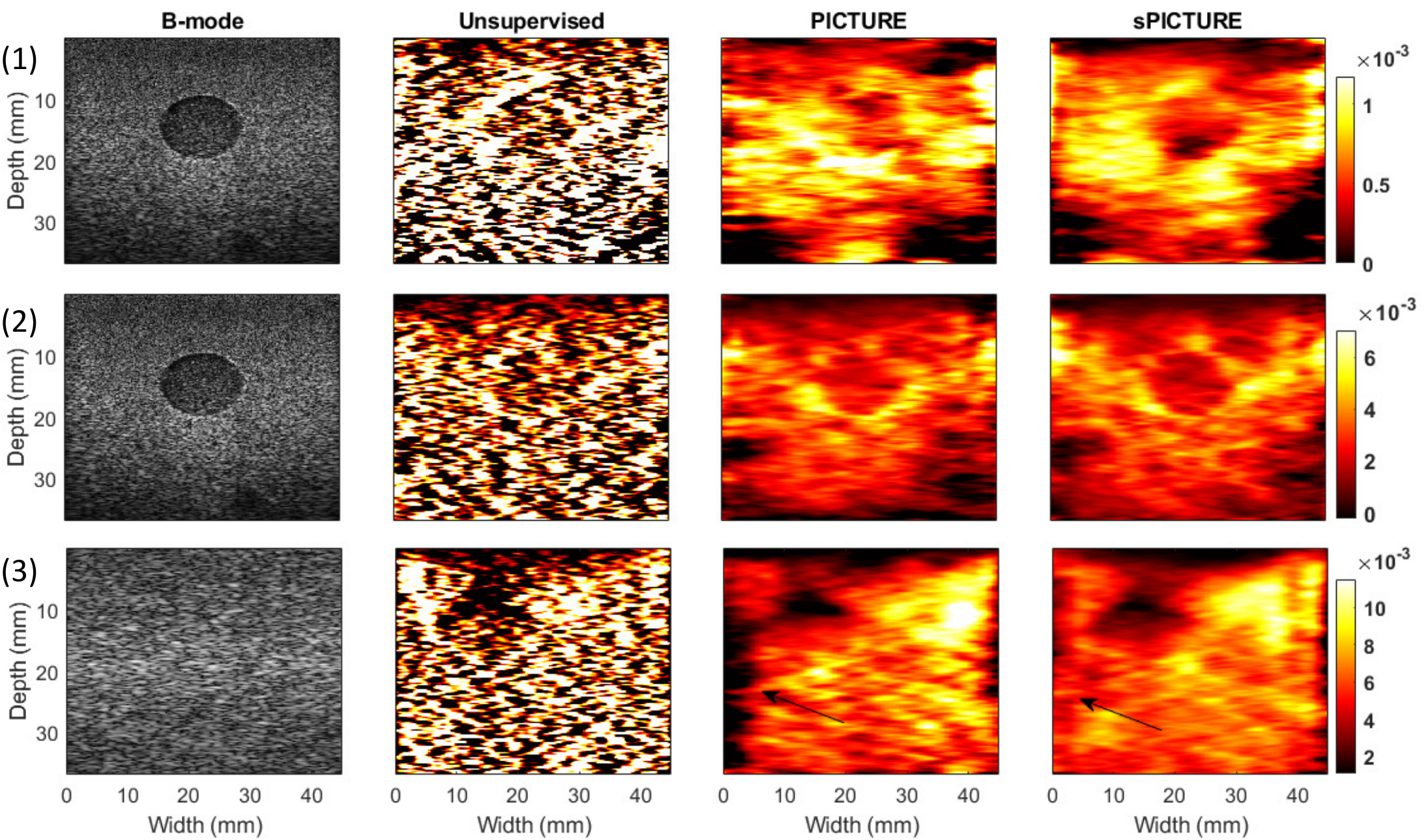}
			\centering
			\caption{Ablation experiment results. In (3), the inclusion is not visible in B-mode image and arrows show that SSL improves the estimation in boundary regions. The samples 1, 2 and 3 are taken from different locations of the tissue-mimicking breast phantom.} 
			\label{fig:ab}
		\end{figure} 
		
		\begin{figure}[t]
			\centering
			\includegraphics[width=0.38\textwidth]{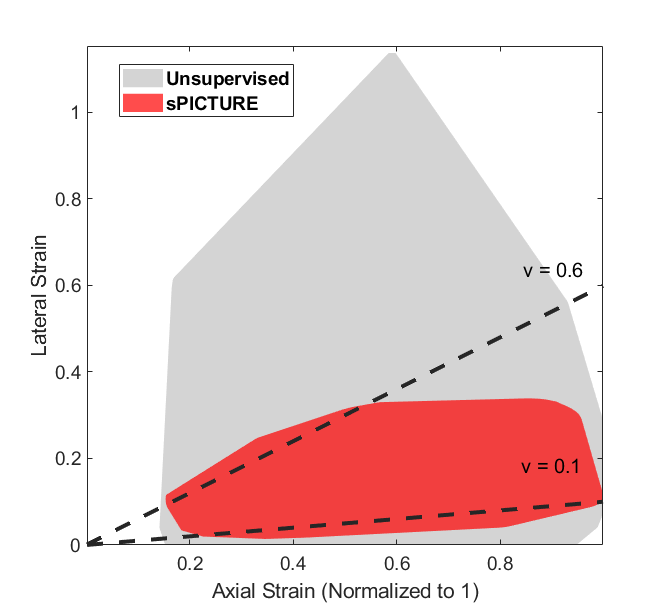}
			\centering
			\caption{Normalized axial strain versus the corresponding lateral strain. The region where the samples of the methods lie for experimental phantom 2 are specified. The regions are obtained from convex hull of strain samples. EPR equals to 0.1 and 0.6 are highlighted by the dashed lines.} 
			\label{fig:poly}
		\end{figure} 
		\begin{figure*}[]
			\centering
			\includegraphics[width=0.999\textwidth]{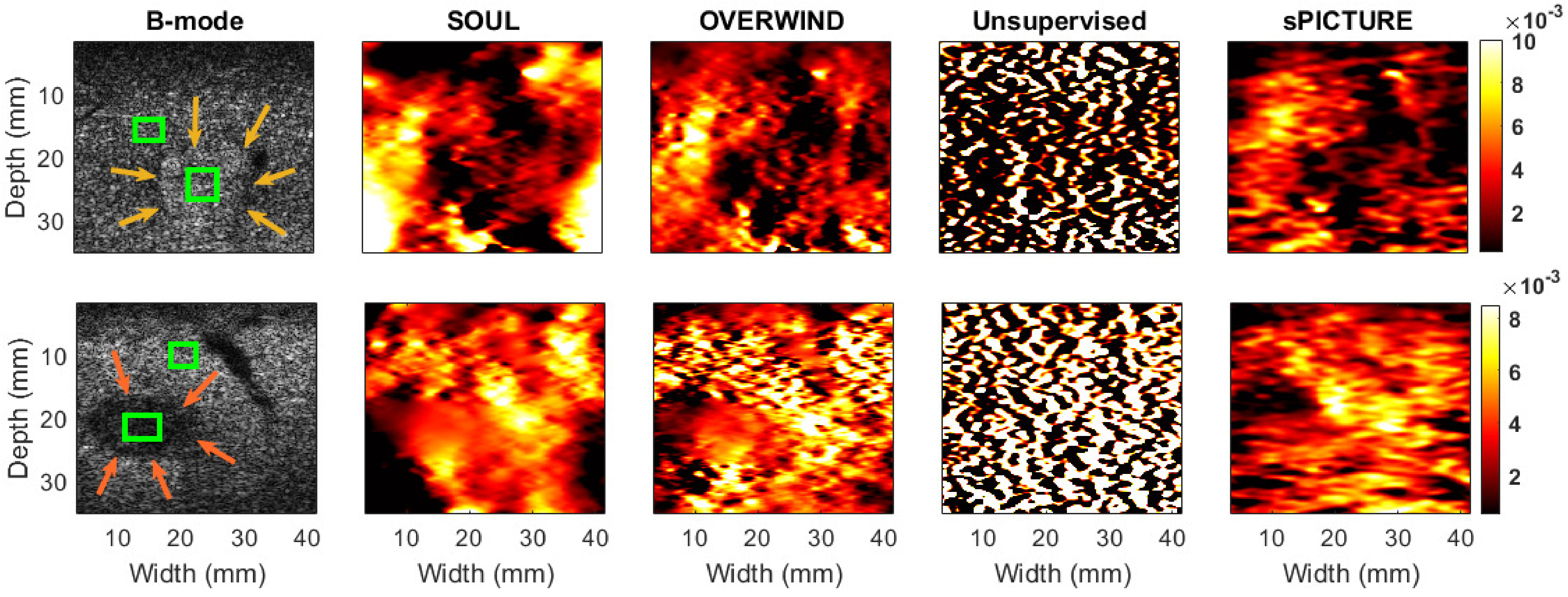}
			\centering
			\caption{The \textit{in vivo} lateral strains obtained by the evaluated methods. The target and background windows for calculation of CNR and SR are marked in the B-mode images. The corresponding axial strains are given in the Supplementary Materials.} 
			\label{fig:invivo}
		\end{figure*}
		\begin{figure}[t]
			\centering
			\includegraphics[width=0.50\textwidth]{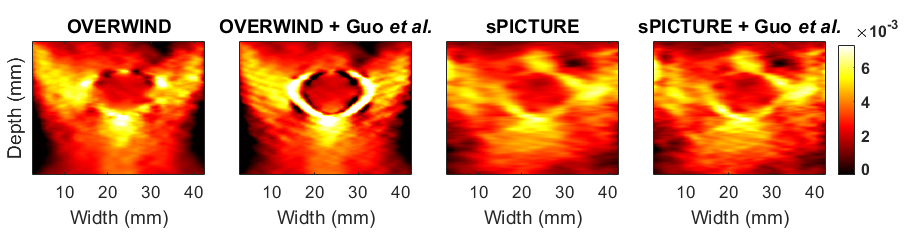}
			\centering
			\caption{Lateral strains of OVERWIND and sPICTURE after applying the method of Guo \textit{et al.} \cite{guo2015pde} to real phantom data (1).}
			\label{fig:pde}
		\end{figure}

		\begin{table*}[tp!]
			\caption{Quantitative results of lateral strains for experimental phantoms and \textit{in vivo} data. Mean and standard deviation ($\pm$) of CNR (higher is better) and SR (lower is better) of lateral strains are reported. The pair marked by asterisk is not statistically significant (\textit{p}-value$>$0.05, using Friedman test).}
			\centering
			\label{tab:phantom}
			\resizebox{0.99\textwidth}{!}{
				\begin{tabular}{@{}ccccccccccc@{}}
					\toprule
					& \multicolumn{2}{c}{Phantom 1}                           & \multicolumn{2}{c}{Phantom 2}                           & \multicolumn{2}{c}{Phantom 3}                           & \multicolumn{2}{c}{\textit{In vivo} 1}    & \multicolumn{2}{c}{\textit{In vivo} 2}    \\
					& CNR           & \cellcolor[HTML]{C0C0C0}SR              & CNR           & \cellcolor[HTML]{C0C0C0}SR              & CNR           & \cellcolor[HTML]{C0C0C0}SR              & CNR & \cellcolor[HTML]{C0C0C0}SR & CNR & \cellcolor[HTML]{C0C0C0}SR \\ \cmidrule(l){2-11} 
					SOUL         & 11.01$\pm$4.52 & \cellcolor[HTML]{C0C0C0}0.483$\pm$0.038 & 3.63$\pm$1.20 & \cellcolor[HTML]{C0C0C0}0.676$\pm$0.081$^\ast$ & 2.33$\pm$0.81 & \cellcolor[HTML]{C0C0C0}0.327$\pm$0.140 &  3.25$\pm$1.07   & \cellcolor[HTML]{C0C0C0} 0.428$\pm$0.137  & 1.26$\pm$ 0.843   & \cellcolor[HTML]{C0C0C0} 1.110$\pm$0.124   \\
					OVERWIND     & 7.21$\pm$1.91 & \cellcolor[HTML]{C0C0C0}0.456$\pm$0.057 & 2.35$\pm$0.68 & \cellcolor[HTML]{C0C0C0}0.676$\pm$0.092$^\ast$ & 3.38$\pm$1.45 & \cellcolor[HTML]{C0C0C0}0.285$\pm$0.155 &1.92$\pm$0.96     & \cellcolor[HTML]{C0C0C0}0.584$\pm$0.150   &  0.84$\pm$0.61   & \cellcolor[HTML]{C0C0C0} 1.048$\pm$0.202  \\ \midrule
					Unsupervised & 2.31$\pm$0.30 & \cellcolor[HTML]{C0C0C0}\textbf{0.454}$\pm$0.061 & 1.02$\pm$0.30 & \cellcolor[HTML]{C0C0C0}0.530$\pm$0.105  & 0.26$\pm$0.18 & \cellcolor[HTML]{C0C0C0}0.677$\pm$0.743 & 0.24$\pm$0.18     & \cellcolor[HTML]{C0C0C0}0.905$\pm$0.159   &  0.50$\pm$0.36   & \cellcolor[HTML]{C0C0C0} 1.125 $\pm$ 0.437  \\
					sPICTURE     & \textbf{11.20}$\pm$2.18 & \cellcolor[HTML]{C0C0C0}0.511$\pm$0.059 & \textbf{9.14}$\pm$2.80 & \cellcolor[HTML]{C0C0C0}\textbf{0.527}$\pm$0.044 & \textbf{7.07}$\pm$1.76 & \cellcolor[HTML]{C0C0C0}\textbf{0.278}$\pm$0.065 & \textbf{7.80}$\pm$2.01     & \cellcolor[HTML]{C0C0C0}\textbf{0.242}$\pm$0.066   &  \textbf{4.34}$\pm$1.39    & \cellcolor[HTML]{C0C0C0} \textbf{0.640}$\pm$0.075  \\ \bottomrule
			\end{tabular}}
		\end{table*}
		
		Unsupervised method has unacceptable results in which the inclusions are not visually detectable while it provides high-quality axial strain images (the axial strain images are shown in the Supplementary Materials) comparable to optimization-based methods. This is an important observation since this shows that the unsupervised loss (composed of data and smoothness losses), which has been used widely in computer vision optical flow estimation,  is not a suitable loss in USE.

		sPICTURE provides high-quality lateral strain images and performs the best in terms of quantitative results among the compared methods. By comparing the unsupervised and sPICTURE results, it can be seen how the added PICTURE regularization and the SSL lead to the improvement of the obtained strain images. The added regularizations convert the unreliable and noisy lateral strains obtained by unsupervised method to the high-quality strain images. It should be mentioned that sPICTURE and unsupervised methods are both trained using the same data and weights for smoothness regularization. Furthermore, sPICTURE obtains substantially higher quality lateral strain images than the compared optimization-based methods (both visually and quantitatively).

		To further analyze the results, the histograms of the EPR of phantom data 3 are depicted in Fig. \ref{fig:hist}. As mentioned earlier, EPR range is similar to the Poisson’s ratio range (0.2 to 0.5, excluding the boundary regions). In PICTURE loss, we penalize EPR values outside the 0.1-0.6 range. The histogram of EPR of unsupervised method covers a wide range of positive and negative values which indicates that many lateral strain values obtained by this method are incorrect. The histogram of the EPR values of OVERWIND and SOUL are more limited than unsupervised method, but they contain values that are negative or higher than 0.8 which is not possible in the phantom. 
		sPICTURE has a more limited range of EPR values but still has some values outside the specified range. The reason is that the proposed PICTURE regularization is only applied during the training phase. Although the proposed PICTURE regularization reduces the range of EPR values, it does not guarantee that all the values fall into the specified range in test time. 
		
		\subsubsection{Ablation Experiment}      
		In order to investigate the impact of PICTURE loss and SSL separately, an ablation experiment is conducted. Fig. \ref{fig:ab} shows the visual comparison of unsupervised method (without PICTURE and SSL), PICTURE (without SSL), and sPICTURE. It is visually clear that both PICTURE and SSL contribute to the improvements obtained by sPICTURE. Without PICTURE, unsupervised method provides noisy and impractical lateral strain images. PICTURE substantially improves the lateral strain image quality and SSL further boosts the quality of the lateral strain image. For instance, in sample (1), the inclusion location can be detected more accurately in sPICTURE compared to PICTURE. Also, the estimation in boundary regions is improved in sPICTURE since it is trained to deal with cropping with SSL. It should be mentioned that SSL without PICTURE was also tested, but it performed inferior to PICTURE. 
		\subsubsection{Experimental Results after applying lateral displacement refinement} \label{sec:pde}
		Lateral displacement refinement of Guo \textit{et al.} \cite{guo2015pde} is applied using the initial displacement obtained by OVERWIND and sPICTURE. It can be observed that this method further improves the lateral displacement estimation, and the initial displacement obtained by sPICTURE provides a high-quality initial value for this method.   
		
		\subsection{\textit{In vivo} Results}
		The \textit{in vivo} lateral strains of two patients with liver cancer are depicted in Fig. \ref{fig:invivo}, quantitative results are reported in Table \ref{tab:phantom}, and their corresponding axial strains are given in the Supplementary Materials. The tumors are more visually detectable in sPICTURE compared to the other methods. Also, quantitative results denote that sPICTURE has the highest CNR and lowest SR values among the compared methods, which confirms the visual analysis.   
		
		\section{Discussion}
		In this paper, a physically inspired regularization to improve the lateral displacement estimation has been proposed. It confines the range of EPR by employing the high-quality axial strain and the known range of EPR values. One limitation of the proposed method is that PICTURE similar to any other form of regularizations is only applied during the training. Even though it limits the range of EPR values in the test time, it does not guarantee that all EPR values be within that range. We observed only a few samples lie outside of the defined range and fixing them during the test time inspired by known operators \cite{maier2019learning} can be an area of future works. 

		It should be mentioned that PICTURE can also have statistical interpretation. The lateral displacement prediction can be viewed as the estimation of a parameter from under sampled and heavily smoothed observations. The conventional methods estimate this parameter in a maximum likelihood (ML) manner without any prior information (only smoothness is considered). However, PICTURE can be viewed as maximum a posteriori (MAP) estimate in which the prior information from compression physics is utilized to find the parameters. Therefore, more reliable lateral displacement can be estimated compared to the conventional methods. To clarify this, the graph of  lateral versus axial strains is depicted in Fig. \ref{fig:poly}. PICTURE enforced the strain samples to lie within $v=0.6$ and $v=0.1$. The areas where the samples of unsupervised method and sPICTURE lie for experimental phantom data 2 are illustrated in the figure. It can be observed that most of the strain samples of sPICTURE lie within the correct range. sPICTURE moved the lateral sample values to the area of the prior knowledge.

		Self-supervision was another regularization that has been used in this work. SSL can prepare the model to deal with corrupted data. In this paper, we applied cropping and added noise. Cropping helps the model deal with boundary regions where finding the correspondence between pre and post-compression images is difficult. Adding noise can also be useful in some scenarios for instance when there is a loss of signal due to high attenuation or there is a cyst where clutter is stronger than the true signal. Applying other forms of transformation such as acoustic noise (reverberation and multiple scattering), inducing decorrelation, and downsampling can be an area of future works.

		In this paper, the performance of lateral displacement refinement method of \cite{guo2015pde} using initial value from sPICTURE and OVERWIND is also investigated. This method is considered as a post-processing method that relies on the initial displacements. We showed that high-quality lateral displacement of sPICTURE can be considered as a good initial value for this method and improves the results of this refinement method.

		Complexity of the training is another issue that should be discussed. We utilized two parallel NVIDIA A100 GPUs with 40 GB of memory each. Even with this size of memory, the maximum batch size that we could train the network was 8. The main reason is that the image sizes are usually large to preserve high-frequency RF data contents and the memory usage is also doubled by the second pass introduced required in SSL. Only the training phase is memory intensive, and inference can be performed with only 5 GB of memory in 140 $ms$ (for an US data of size $2048\times256$) similar to MPWC-Net++.
		
		\section{Conclusion}
		In this paper, we proposed PICTURE to improve the lateral strain images in USE using physically inspired priors. We further improved the method in sPICTURE by adding the self-supervision to the method. The effectiveness of the proposed method is validated using simulation, experimental phantom, and \textit{in vivo} data.
		
		\section{Acknowledgment}
		This work is Supported by Natural Sciences and Engineering Research Council of Canada (NSERC) Discovery Grant. The Alpinion ultrasound machine was partly funded by Dr. Louis G. Johnson Foundation.
		The authors would like to thank Drs. E. Boctor, M. Choti and G. Hager for providing us with the \textit{in vivo} patients data from Johns Hopkins Hospital.

		\bibliographystyle{IEEEtran}
		\bibliography{IEEEfull}
		
	\end{document}